\documentclass{aa}
\usepackage[breaklinks=true,colorlinks,citecolor=blue]{hyperref}
\usepackage{amsmath}
\usepackage{units}
\usepackage{gensymb}
\usepackage{txfonts}

\newcommand{\modified}[1]{#1}  

\graphicspath{{./}{figures/}}

\begin{document}
\title{Rotation in Stellar Evolution: Probing the Influence on Population Synthesis in High-Redshift Galaxies}	
\titlerunning{Rotation in population synthesis}
\author{Weijia Sun$^1$\thanks{E-mail: wsun@aip.de}} 

\authorrunning{Weijia Sun}

\institute{$^1$ Leibniz-Institut für Astrophysik Potsdam (AIP), An der Sternwarte 16, 14482 Potsdam, Germany}
        
\date{Received ; accepted }
  
\abstract{
Stellar population synthesis (SPS) is essential for understanding galaxy formation and evolution. However, the recent discovery of rotation-driven phenomena in star clusters warrants a review of uncertainties in SPS models caused by overlooked factors, including stellar rotation. In this study, we investigate the impact of rotation on SPS specifically using the PARSEC V2.0 rotation model and its implications for high redshift galaxies with the JWST. Rotation enhances the ultraviolet (UV) flux for up to $\sim\unit[400]{Myr}$ after the starburst, with the slope of UV increasing as the population gets faster rotating and more metal-poor. Using the \texttt{Prospector} tool, we construct simulated galaxies and deduce their properties associated with dust and star formation. Our results suggest that rapid rotation models result in a gradual UV slope up to \unit[0.1]{dex} higher and an approximately 50\% increase in dust attenuation for identical wide-band spectral energy distributions. Furthermore, we investigate biases if the stellar population should be characterized by rapid rotation and demonstrate that accurate estimation can be achieved for rotation rates up to $\omega_\text{i}=0.6$. Accounting for the bias in the case of rapid rotation aligns specific star formation rates more closely with predictions from theoretical models. Notably, this also implies a slightly higher level of dust attenuation than previously anticipated, while still allowing for a `dust-free' interpretation of the galaxy. The impact of rapid rotation SPS models on the rest-UV luminosity function is found to be minimal. Overall, our findings have potentially important implications for comprehending dust attenuation and mass assembly history in the high-redshift Universe. 
}

\maketitle
  
\section{Introduction}
\label{sec:intro}
Stellar population synthesis (SPS) lies at the heart of our modern understanding of galaxy formation and evolution \citep{1980FCPh....5..287T}. Initially pioneered by \citet{1968ApJ...151..547T, 1971ApJS...22..445S}, SPS integrates stellar evolution, stellar atmospheric models, and star formation theory, becoming indispensable for estimating parameters such as stellar mass-to-light ratios, star formation rates, stellar ages, and metallicities \citep[e.g., ][]{2000AJ....119.1645T, 2009ApJS..184..100L, 2011MNRAS.415..709S, 2013ApJ...767...50M}. Despite its influential role, SPS is challenged by uncertainties stemming from the assumptions and constraints of its models \citep{2013ARA&A..51..393C}.

Uncertainties in stellar evolution models, stellar spectral libraries, and initial mass functions introduce complications in the construction of simple stellar populations (SSPs), the cornerstone of SPS. These SSPs represent the time-dependent evolution of coeval and mono-metallic stellar populations' spectral energy distributions (SEDs), with star clusters serving as prototypes. The examination of these stellar populations, as illustrated by tests using their color-magnitude diagrams \citep[CMDs, e.g.,][]{2003MNRAS.344.1000B, 2010MNRAS.404.1639V}, forms an essential bridge between theoretical stellar models and practical SPS application. 

However, observed phenomena such as the extended main-sequence turn-off (eMSTO) and split main sequences within young and intermediate-age star clusters ($t \leqslant \unit[2]{Gyr}$) challenge the traditional assumption that star clusters are authentic manifestations of SSPs \citep{2007MNRAS.379..151M, 2011ApJ...737....3G, 2015MNRAS.453.2637D, 2017ApJ...844..119L}. Notably, the MSTO and upper main sequence (MS) regions of these clusters exhibit significant broadening, while the lower MS region remains compact---a conflict against traditional SSP assumptions. Moreover, these features are pervasive in young and intermediate-age clusters within the Magellanic Clouds \citep{2009A&A...497..755M, 2010MNRAS.403.1156R, 2013MNRAS.431.3501G} and a substantial portion of open clusters in the Milky Way \citep{2018ApJ...869..139C}. The spectroscopic observations of these clusters confirm that stellar rotation significantly influences the morphology of their color-magnitude diagrams \citep{2017ApJ...846L...1D, 2018MNRAS.480.3739B, 2019ApJ...876..113S, 2019ApJ...883..182S, 2020MNRAS.492.2177K}. The bimodal rotational distribution, predominantly applicable to stars exceeding $\sim\unit[2.5]{M_\odot}$ in the field \citep{2010ApJ...722..605H, 2012A&A...537A.120Z, 2021ApJ...921..145S}, challenges the adequacy of previous SPS models that neglected this component, presenting a significant uncertainty in current SPS applications. Therefore, the aspiration to conceive a more accurate and encompassing SPS model calls for a thorough investigation into the effects of stellar rotations and their consequent manifestations.

In this work, we investigate the impact of stellar rotation on SPS and explore its variation with different rotation rates. Building upon the work of \citet{2017ApJ...838..159C}, where the authors extended SPS models to include contributions from very massive stars ($>\unit[100]{M_\odot}$) and found that fast-rotating stars are crucial in sustaining the production of ionizing photons beyond a few Myr following a starburst ($\leqslant\unit[5]{Myr}$), we shift our focus towards the subsequent evolutionary phase ($>\unit[20]{Myr}$) predominantly contributed by less massive stars ($\sim \unit[10]{M_\odot}$). Furthermore, we extend these findings within the SPS framework to high-redshift ($z\sim 10$) galaxies observable with the James Webb Space Telescope (JWST). JWST, with its 6.5 m aperture and infrared capabilities \citep{2023PASP..135d8001R}, has revolutionized the study of high-redshift galaxies in the early Universe. In its first year after launch, JWST has already identified numerous distant galaxy candidates up to $z \approx 15$ \citep{2023MNRAS.518.6011D}, providing insights into their minimal dust content \citep{2022ApJ...940L..14N, 2022ApJ...938L..15C, 2023ApJ...942L..27S, 2023MNRAS.518.4755A} and direct metallicity measurements at $z \approx 8$\citep{2023MNRAS.518..425C, 2023NatAs...7..622C}. These groundbreaking observations, characterized by unexpectedly bright galaxies and high stellar masses, challenge current models of galaxy formation. Using the \texttt{Prospector} tool \citep{2017ApJ...837..170L}, we investigate the consequences of rapid rotation in the interpretation of SEDs of the galaxies \modified{and examine potential biases in their properties if galaxies are composed of non-rotating stellar populations}.

This article is organized as follows. In Section \ref{sec:model}, we provide a brief overview of the stellar evolutionary models with rotation and the SPS models used in this work. Section \ref{sec:result} explores the influence of rotation on spectroscopic and photometric properties of stellar population observables, examining its dependency on stellar mass and metallicity. In Section \ref{sec:cluster}, we present a comparison of the models with observed integrated spectra of star clusters of various ages and metallicities. Section \ref{sec:discussion} discusses the implications of our models for galaxies observed in the high-redshift universe based on JWST multi-band observations. Finally, Section \ref{sec:summary} summarizes our conclusions.

\section{Models}
\label{sec:model}
In this section, we provide an overview of the stellar evolutionary models with rotation and the SPS models used in this study.

\subsection{Stellar evolution}
\label{sec:sp}
We adopt the stellar evolutionary tracks from the PAdova and tRieste Stellar Evolutionary Code (PARSEC) V2.0 \citep{2022A&A...665A.126N}. These models have undergone recent updates since achieved in \citet{2012MNRAS.427..127B}, including improvements in the treatment of convective zones, mass loss, nuclear reaction networks, and other physical input parameters. The PARSEC V2.0 models \modified{, first implemented in \citet{2019MNRAS.485.4641C},} cover a range of initial mass from $\unit[0.09]{M_\odot}$ to $\unit[14]{M_\odot}$ and six sets of initial metallicity from $Z = 0.004$ to $Z = 0.017$ (solar-scaled chemical mixtures $Z_\odot = 0.01524$).

In PARSEC V2.0, the basic quantity describing the effect of rotation in the stellar structure is the initial angular rotation rate, $\omega$. This is defined as
\begin{equation}
	\omega = \frac{\Omega}{\Omega_\mathrm{c}}
\end{equation}

where $\Omega$ is the angular velocity and $\Omega_\mathrm{c}$ is the critical angular velocity $\Omega_\mathrm{c}=\left(\frac{2}{3}\right)^{3/2}\sqrt{\frac{GM}{R^3_\mathrm{pol}}}$, with the gravitational constant ($G$), the mass ($M$) enclosed by the polar radius ($R_\mathrm{pol}$). The PARSEC models include seven angular rotation rates, ranging from non-rotating models ($\omega_\mathrm{i} = 0$) to models with intermediate rotation ($\omega_\text{i} = 0.3, 0.6, 0.8$), and models with high rotation rates ($\omega_\mathrm{i} = 0.9, 0.95, 0.99$) close to the critical break-up rotational velocity.

For low-mass stars ($\lesssim\unit[1]{M_\odot}$), rotation has minimal influence on their evolution \citep[e.g.,][]{2014ApJS..211...24M}. The rotation velocities of these stars are significantly lower compared to intermediate and high-mass stars, except during the pre-main-sequence phase \citep{2021ApJ...923..177S}. This break is attributed to angular momentum loss through magnetized winds and sporadic mass ejections from stars with deep surface convection zones \citep{1962AnAp...25...18S}. Therefore, the PARSEC models assume non-rotating behavior for stars with a mass lower than $\sim\unit[1]{M_\odot}$, as these stars do not exhibit significant rotation-induced effects. A smooth transition is implemented between non-rotating and fast-rotating models for stars with masses between these thresholds.

The influence of rotation on stellar evolutionary tracks can be summarised into two main aspects: gravity-darkening and rotational mixing. Gravity-darkening refers to the effect of rotation on the distribution of surface brightness across a star's surface. As a star rotates, the centrifugal force reduces the stellar surface gravity along the equator, causing a decrease in both the local surface effective temperature and the core pressure \citep{1968ApJ...151..203F}. On the other hand, rotational mixing describes the process by which rotation can induce extra mixing within a star \citep{2000ARA&A..38..143M}. As a star rotates, fresh fuel at the equator is lifted towards the surface, while material at the poles sinks towards the core. This mixing of materials with different chemical compositions can impact the star's overall evolution and can extend its main-sequence lifetime. While this prolonging happens for all stellar masses over $\sim\unit[1]{M_\odot}$, the meridional circulation that carries the fresh fuel to the central core works much more efficiently in the intermediate-mass ($\geqslant\unit[1.8]{M_\odot}$) and more massive stars \citep[see Figure 3 in][]{2022A&A...665A.126N}. These phenomena lead to a larger core mass, making the luminosity remain higher during the post-main-sequence evolutionary phases. 

It is worth noting that, in the case of rotating stars, the effective temperature ($T_\mathrm{eff}$) is not a constant value across the star's surface. Instead, in PARSEC model, it represents an average value over the star's isobaric surface and specifically corresponds to the temperature that a non-rotating star with the same ``volumetric radius'' would need to have to produce the same total luminosity. The volumetric radius is defined as the radius of a sphere that has the same volume as a rotating star. Conversely, the local effective temperature that characterizes different points on the surface of a rotating star varies along the co-latitude angle, becoming cooler towards the equator. This phenomenon can be described by von Zeipel's theorem \citep{1924MNRAS..84..665V}.

\begin{figure*}[ht!]
\centering
\includegraphics{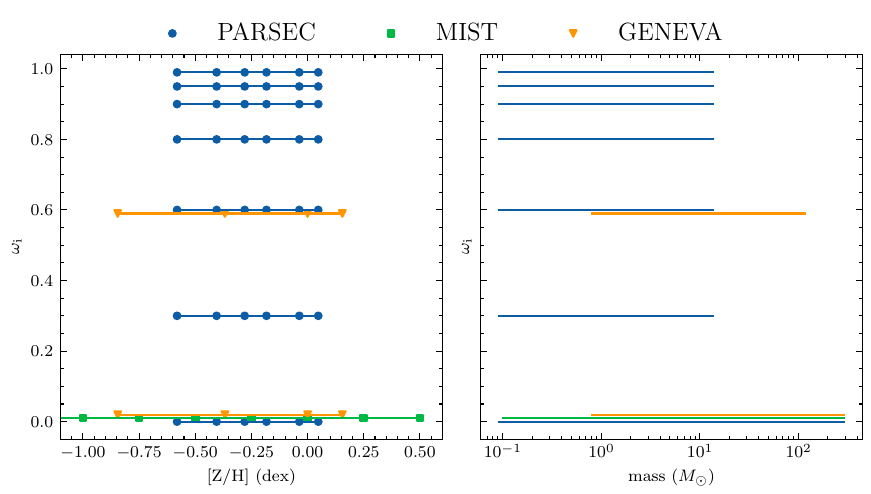}
\caption{Coverage of the ($\omega_\mathrm{i}$--[Z/H], left) and the ($\omega_\mathrm{i}$--mass, right) plane by the models investigated in this work (PARSEC: blue circle, MIST: \modified{green} square, GENEVA: \modified{orange} triangle). The MIST models extend to lower metallicity but only values of $\mathrm{[Z/H]} \geqslant \unit[-1]{dex}$ are shown on the right panel. The stellar mass ranges covered by these models are displayed on the right panel using a logarithmic scale and remain consistent across the [Z/H] grid within each model. Slight adjustments have been made to the $\omega_\mathrm{i}$ values of the MIST and GENEVA models for better visualization. \label{fig:modelgrid}}
\end{figure*}

Although the PARSEC model provides a state-of-the-art description of rotation in stellar evolution, it has limited applicability as it is designed for a small parameter space in terms of stellar mass and metallicity (see Fig.~\ref{fig:modelgrid}). To overcome this limitation and ensure a broader coverage, we have also studied two other models: the GENEVA stellar models \citep{2012A&A...537A.146E, 2012A&A...542A..29G, 2013MNRAS.433.1114Y, 2014A&A...566A..21G} and the Mesa Isochrones and Stellar Tracks \citep[MIST,][]{2016ApJ...823..102C} models.

The GENEVA models (green in Fig.~\ref{fig:modelgrid}) take into account the changes in the internal structure of a rotating star, including the mixing of elements, the redistribution of angular momentum, magnetic fields, and mass loss \citep{2012A&A...537A.146E, 2012A&A...542A..29G, 2013MNRAS.433.1114Y, 2014A&A...566A..21G}. These models cover a wide range of stellar masses, from 0.8 to $\unit[120]{M_\odot}$, and metallicities of $Z = 0.002, 0.006, 0.014, 0.02$. Two rotation rates are considered: non-rotating models ($\omega = 0$) and models with a moderate rotation rate ($\omega = 0.568$, corresponding to a ratio of rotational velocity to critical rotational velocity, $v/v_\mathrm{crit} = 0.4$).

As a reference non-rotating evolutionary model, we also include the MIST models (cyan in Fig.~\ref{fig:modelgrid}). These models self-consistently and continuously evolve from the pre-main sequence to the end of hydrogen burning, the white dwarf cooling sequence, or the end of carbon burning, depending on the initial mass, depending on the initial mass. The MIST models cover a broad range of masses, from $\unit[0.1]{M_\odot}$ to $\unit[300]{M_\odot}$, and have solar-scaled abundance ratios.

Note in the GEVENA model, the most massive models ($\unit[120-500]{M_\odot}$) have evolved without the insurance of angular momentum conservation. To ensure continuity in the modeling, we followed a similar practice and implemented the heavy end ($M \geqslant\unit[14]{M_\odot}$) of the PARSEC models with those of the same metallicity from the non-rotating PARSEC 1.2S models \citep{2012MNRAS.427..127B}, which share same evolution before the end of MS \citep{2022A&A...665A.126N}.

\subsection{Stellar population synthesis}
\label{sec:sps}

We utilize the Flexible Stellar Population Synthesis package \citep{2009ApJ...699..486C,2010ApJ...712..833C,2010ApJ...708...58C,2010ascl.soft10043C} to generate a series of stellar populations using a specific model with different rotation rates as outlined in the previous section. The \texttt{FSPS} code is accessed through the \texttt{Python-FSPS} \citep{2014zndo.....12157F} bindings.

We adopt the stellar spectral library from the MILES empirical library \citep{2006MNRAS.371..703S, 2011A&A...532A..95F} and the dust emission libraries from \citet{2007ApJ...657..810D}. An initial mass function (IMF) from \citet{2001MNRAS.322..231K} was adopted. Stars older than 10 Myr are attenuated following the \citet{2000ApJ...533..682C} attenuation curve, which is applied to all starlight equally.

There are other inputs (e.g., Wolf--Rayet, binaries) that may also influence the properties of synthetic SSP as a factor of rotation. For instance, The ratio of red to blue supergiant could be altered as rotation changes lower the opacity in the radiative envelope and increase the luminosity. Also, the mixing to the surface of hydrogen-burning products caused by rotation will affect the number and type of WR stars as well as their signatures as a function of time and metallicity in integrated spectra \citep{2007ApJ...663..995V, 2009ApJ...699..486C}. However, for this work, the contribution from these massive components in the mass spectrum would quickly decay within the first $\unit[10]{Myr}$ of the stellar evolution. \citet{2007ApJ...663..995V} found the ratio of Wolf--Rayet stars over O-type stars is significant only within the first \unit[7]{Myr} since the stellar formation, even with the prolong of its lifetime due to rotational mixing. Such arguments would be the same correct for most of the super giants. And by the age of $\sim\unit[30]{Myr}$ (see Section~\ref{sec:mass}), the integrated light would be dominated by stars within the mass range of the PARSEC models. Therefore, the lack of these elements will not hinder our investigation of the role of stellar rotation in the stellar population under study.

\section{Results}
\label{sec:result}
The previous sections have laid the groundwork for the stellar evolution model and the SPS tool we employed. In this section, we delve into the critical aspects of derived SPS properties. Specifically, we analyze the influence of rotation on key stellar population observables and explore how they vary with metallicity. We also examine the mass contribution to the stellar population and its role in characterizing the overall composition. Furthermore, we assess the consistency of these results across different models. 

\subsection{Dependence on stellar mass}
\label{sec:mass}
\begin{figure*}[ht!]
\centering
\includegraphics{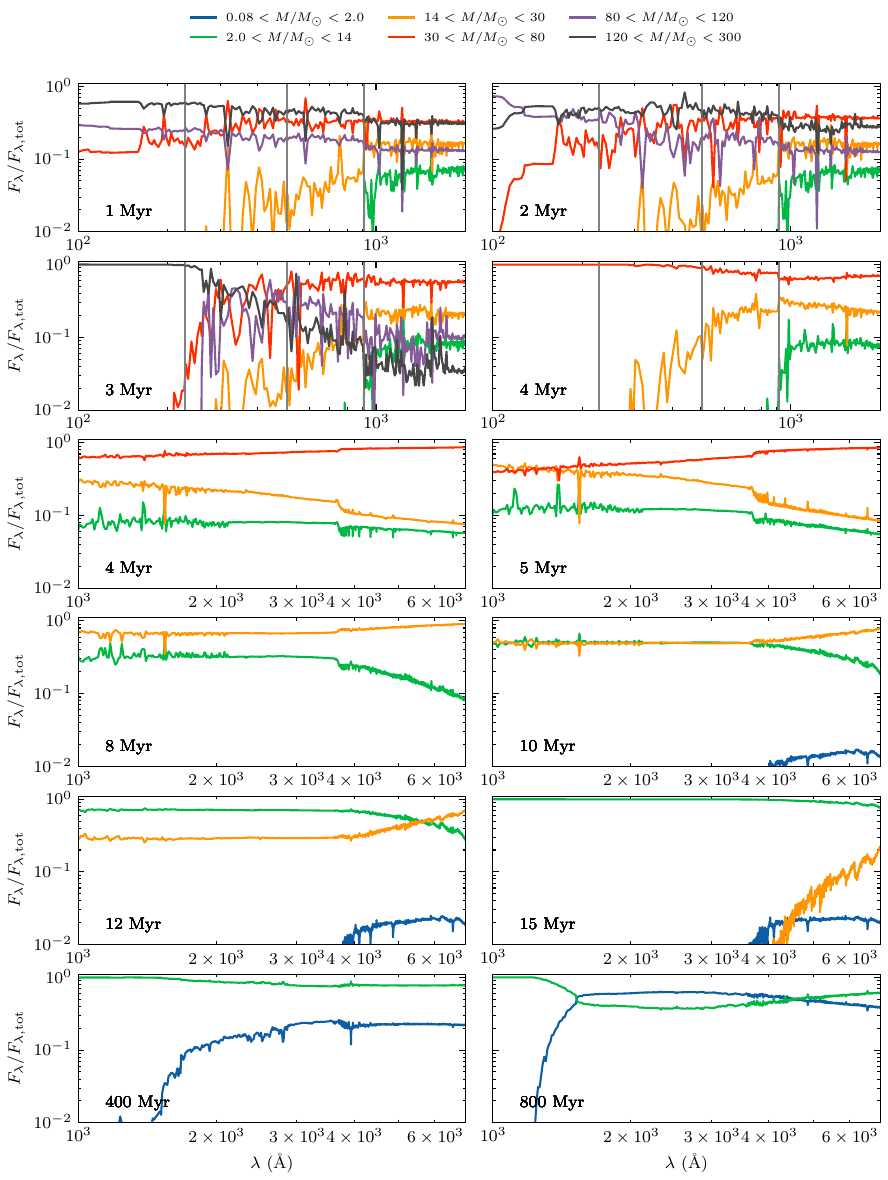}
\caption{Fractional flux contribution to the total UV or optical flux in different mass ranges for a single-burst population with the MIST stellar evolution library, without rotation. UV light is found to be significant within the first few million years, as the contribution from massive stars quickly decreases. Intermediate-mass stars ($\unit[2]{M_\odot}\leqslant M \leqslant \unit[30]{M_\odot}$) dominate the photons across the optical wavelength from approximately $\unit[5]{Myr}$ until around $\unit[800]{Myr}$. In the top two rows, the gray vertical lines represent the wavelengths below which photons can ionize hydrogen, singly ionize helium, and doubly ionize helium ($\unit[912]{\AA}$, $\unit[504]{\AA}$, and $\unit[228]{\AA}$, respectively. \label{fig:specbymass}}
\end{figure*}

One critical aspect of SPS is to understand the flux contribution of different stellar masses. To investigate this, we computed the fractional flux contribution to the total UV and optical flux in different mass ranges for a single-burst stellar population using the MIST stellar evolution library. \modified{It is noteworthy, however, that while intermediate-mass stars significantly influence the infrared bands, in this study, our emphasis on the UV and optical spectrum is driven by the higher flux density in these ranges, which is more pivotal for analyzing the characteristics of younger stellar populations.} Fig.~\ref{fig:specbymass} presents these results, showing the fractional contribution at different ages. The upper limit of the IMF was set to $\unit[300]{M_\odot}$. The top two rows of subfigures correspond to ages ranging from $\unit[1-4]{Myr}$, while the middle three rows show ages ranging from $\unit[4-15]{Myr}$. The bottom row represents ages ranging from $\unit[400-800]{Myr}$. In the top rows, which cover wavelengths from $\unit[200-2000]{\AA}$, the most massive stars are found to contribute significantly to the UV flux within the first few million years. Conversely, the middle rows, which cover optical flux from $\unit[1000-7000]{\AA}$, illustrate that less massive stars ($\unit[2]{M_\odot}\leqslant M/ \leqslant \unit[30]{M_\odot}$) dominate the optical spectra after approximately $\unit[5]{Myr}$. Notably, it is only at around $\unit[800]{Myr}$, when the lowest mass stars ($\unit[0.08]{M_\odot}\leqslant M \leqslant \unit[2]{M_\odot}$) whose evolution is not influenced by rotation, that they determine the major features of the integrated light of the entire population.

Investigating the flux contribution of different mass ranges is crucial for our study, particularly due to the limited coverage of the finest recipe for stellar rotation in the PARSEC model, which extends only up to a mass of $\unit[14]{M_\odot}$. Fig.~\ref{fig:specbymass} illustrates the varying contributions of different mass ranges to the total UV and optical flux, highlighting their distinct evolution with age. Notably, within the mass range covered by the PARSEC model ($M\leqslant\unit[14]{M_\odot}$), these contributions exhibit comparable significance to their more massive counterparts at approximately $\unit[10]{Myr}$ and subsequently become the primary component. It is important to acknowledge that the precise timing of this transition may be influenced by other factors, such as metallicity (see Section \ref{sec:metallicity}). However, it is worth noting that the SPS based on the PARSEC model remains valid for any SSP older than approximately $\unit[10]{Myr}$, thereby establishing the importance of considering the impact of stellar rotation in population synthesis analyses. A similar applicability can be established for GENEVA models, ranging from approximately $\unit[4]{Myr}$ to $\unit[400]{Myr}$.

\subsection{Dependence on rotation rate}
\label{sec:rotation} 
\begin{figure*}[ht!]
\centering
\includegraphics{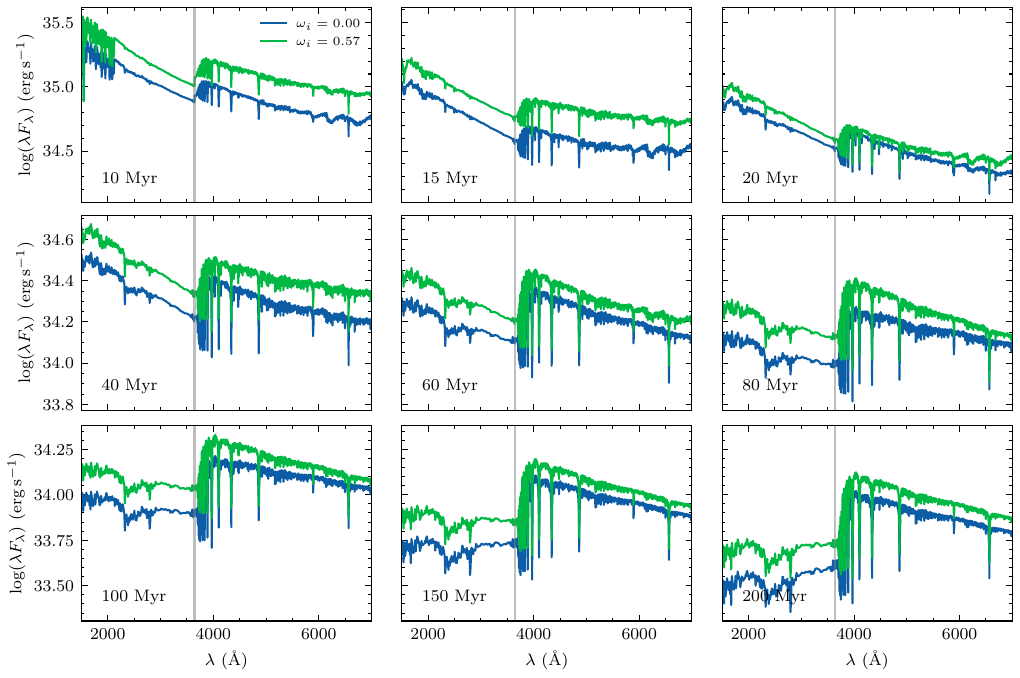}
\caption{Time evolution of SED predictions for an SSP at solar metallicity ([Fe/H]=0) from GENEVA models. The spectra have been normalized to an initial total mass of $\unit[1]{M_\odot}$. Different colors represent different initial rotation rates, denoted as $\omega_i$. The gray vertical lines highlight the wavelength of the Balmer jump at $\unit[3645]{\AA}$. \label{fig:geneva_spec_rot}}
\end{figure*}

\begin{figure*}[ht!]
\centering
\includegraphics{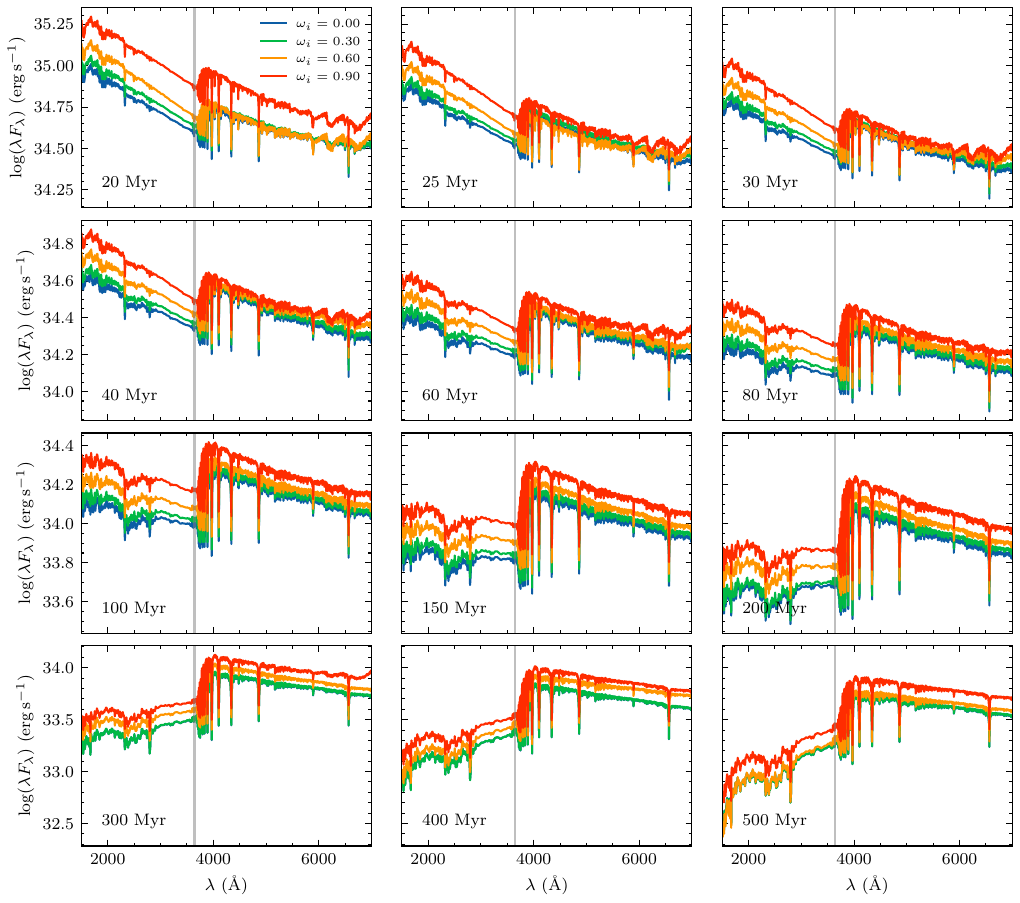}
\caption{Same as Figure \ref{fig:geneva_spec_rot} but for PARSEC models. \label{fig:parsec_spec_rot}}
\end{figure*}

We will begin by examining single rotation rates ($\omega_\mathrm{i}$) in the context of an SSP. Fig.~\ref{fig:geneva_spec_rot} and \ref{fig:parsec_spec_rot} present the time evolution of SED predictions for an SSP at solar metallicity ([Fe/H]=0) from GENEVA and PARSEC models, respectively. In both figures, the spectra have been normalized to an initial total mass of $\unit[1]{M_\odot}$. Non-rotating models are represented by the color blue, while other colors (green, orange, and red) represent rotating models with varying non-zero initial rotation rates. Rotation leads to an overall increase in flux across a range of wavelengths compared to non-rotating counterparts. This flux enhancement is a consequence of the additional mixing induced by rotation, which extends the lifetimes and increases the luminosities of stars. As depicted in Fig.~\ref{fig:parsec_spec_rot}, the influence of rotation is particularly pronounced with faster initial rotation rates. In the bottom-left panel of Fig.~\ref{fig:parsec_spec_rot}, the SSPs with $\omega_i = 0.00$ and 0.30 largely converge, indicating that the influence of intermediate rotation rates diminishes as the main flux contribution shifts towards low-mass stars, which are not significantly affected by additional rotational evolution prescriptions. However, one exception is the case of $\omega_i = 0.90$ (represented by red), which remains prominent even at an age of around $\unit[500]{Myr}$.

As for the spectral features, there are no peculiar characteristics specifically linked to rapid rotation, as depicted in Fig.~ \ref{fig:geneva_spec_rot} and \ref{fig:parsec_spec_rot}. This is expected behavior in the context of SPS, in which models built upon empirical libraries without any special treatment of rotation do not include rotation-only spectral features \citep[e.g., Be-star;][]{2004MNRAS.350..189T}. \modified{Moreover, it is important to note that the consideration of rotational-broadened profiles, commonly observed in individual rapidly rotating stars, takes on a different perspective in SPS modeling. While rotational broadening in a single star results in a spectrum that varies with the observer's orientation --- due to the different line-of-sight velocities across the star's disk --- this effect is not directly applicable in SPS. This is because SPS focuses on the total flux emitted by an entire stellar population, not just individual stars. The cumulative flux from all stars in the population, regardless of the observer's perspective, is what determines the observed spectral properties. Therefore, the rotational broadening spectral features observed in rapidly rotating stars are averaged out in the aggregated flux of the whole population, rendering it a non-critical factor in SPS modeling.}

\begin{figure*}[ht!]
\centering
\includegraphics{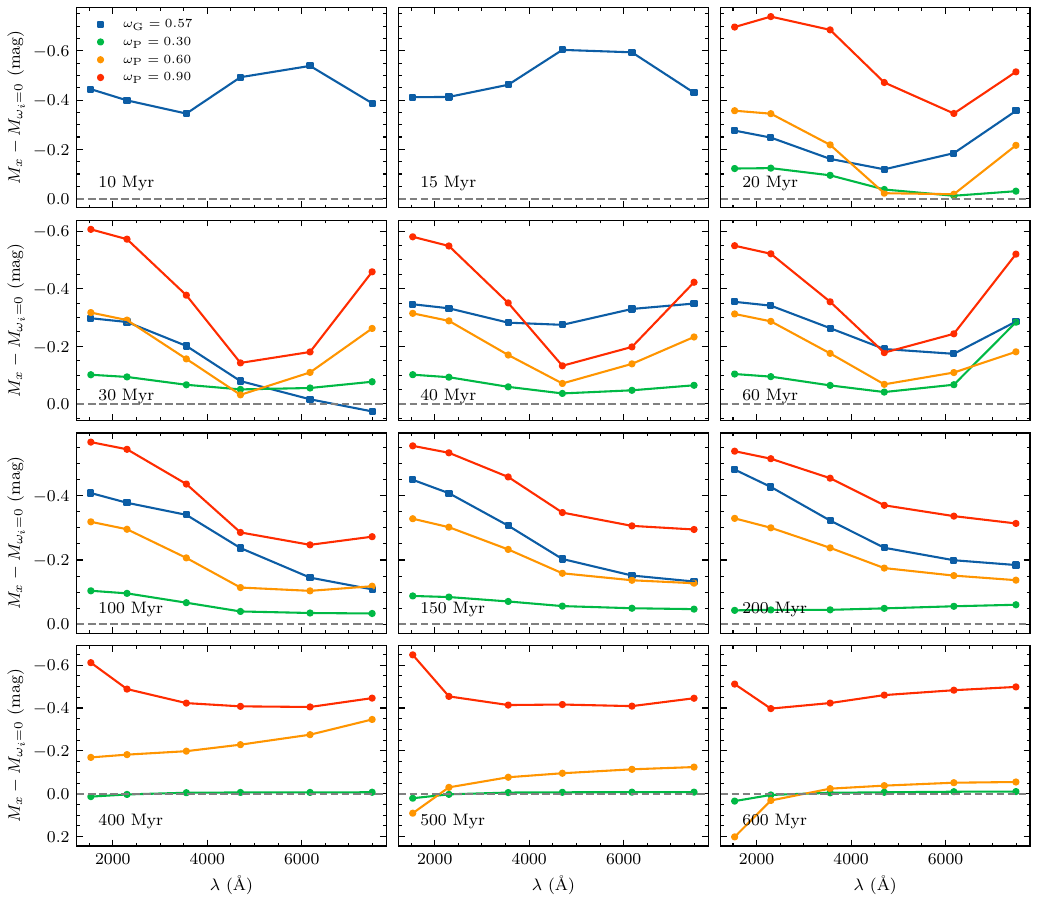}
\caption{Time evolution of photometry variation ($M_x - M_{\omega_i=0}$) caused by rotation predicted by GENEVA and PARSEC models at solar metallicity ([Fe/H]=0). The SEDs are convolved into six passbands: GALEX NUV, NUV, and SDSS $u$, $g$, $r$, and $i$ filters. GENEVA models are shown for populations ages younger than $\unit[300]{Myr}$, while PARSEC models are shown for populations older than $\unit[20]{Myr}$. \label{fig:phot_rot}}
\end{figure*}

To further explore the impact of rotation on stellar populations, we also investigate the photometric properties of rotating SSPs. Fig.~\ref{fig:phot_rot} illustrates the photometric variations caused by rotation as predicted by GENEVA and PARSEC models at solar metallicity. These photometric variations are obtained by convolving the SEDs into six passbands: Galaxy Evolution Explorer (GALEX) far-ultraviolet (FUV) and near-ultraviolet (NUV) bands \citep{2005ApJ...619L...1M, 2005ApJ...619L...7M}, as well as the Sloan Digital Sky Survey (SDSS) $u$, $g$, $r$ at \unit[3556]{\AA}, \unit[4702]{\AA}, \unit[6175]{\AA}, and \unit[7490]{\AA} filters, respectively \citep{1996AJ....111.1748F, 1998AJ....116.3040G, 2002AJ....123.2121S}. The photometric variations illustrate how stellar rotation affects the SEDs by comparing them to the non-rotating model predictions. This allows for a direct comparison of the GENEVA and PARSEC models, taking into account the differences in mass ranges represented by the two sets of models (as shown in Fig.~\ref{fig:modelgrid}).

Fig.~\ref{fig:phot_rot} presents the time evolution of photometry variations ($\Delta M_x = M_x - M_{\omega_i=0}$) caused by rotation as predicted by GENEVA and PARSEC models at [Fe/H]=0, where $x$ represents different stellar evolution models. As illustrated in Fig.~\ref{fig:geneva_spec_rot} and \ref{fig:parsec_spec_rot}, there is a noticeable flux boost for a moderate initial rotation rate of around $\omega_\text{i}=0.60$ that persists up to an age of \unit[400]{Myr}. Notably, the rotating GENEVA model ($\omega_{i, \mathrm{G}} = 0.57$) follows a similar trend to the PARSEC model with the same initial rotation rate ($\omega_{i, \mathrm{P}} = 0.60$). Starting from around \unit[10]{Myr}, the SEDs of intermediate rotating models consistently exhibit a uniform brightness increase of approximately $\unit[0.2-0.4]{mag}$ from the FUV to the optical range. The enhancement is particularly pronounced in the blue end of the spectrum and persists until $t=\unit[400]{Myr}$. However, after approximately $\sim\unit[500]{Myr}$, the trends begin to invert, and the UV flux is depressed by rapid rotation. \modified{For rapid-rotating population ($\omega_\text{i}=0.90$), the inversion happens at around \unit[700]{Myr}.} This UV dimming is linked to gravity-darkening effects, where the reduction in local effective gravity, due to the higher centrifugal force, leads to lower average $T_\mathrm{eff}^4$ and thus a cooler state for the stars during their MS phase. 

Similar to Fig.~\ref{fig:parsec_spec_rot}, the photometric variation induced by rotation is positively correlated with the initial rotation rate. Specifically, the difference in photometry between the rotating models and non-rotating models at $\omega_\mathrm{i}=0.30$ generally exhibits a milder extent of around \unit[0.2]{mag} compared to that of $\omega_\mathrm{i}=0.60$. After a time of approximately \unit[500]{Myr}, the difference in photometry between the $\omega_\mathrm{i}=0.30$ and $\omega_\mathrm{i}=0$ models becomes negligible. However, for the extreme case of $\omega_\mathrm{i}=0.90$, the difference in photometry compared to the $\omega_\mathrm{i}=0.30$ model could be larger than \unit[0.6]{mag} in the UV regime.

\subsection{Dependence on metallicity}
\label{sec:metallicity}
Stellar rotation has been observed to have a significant impact on SPS, with its effects varying depending on the metallicity of the stars. Massive stars have been noted to exhibit higher intrinsic rotation speeds compared to their less massive counterparts \citep{2015A&A...580A..92R}. This difference becomes even more pronounced at lower metallicities, where diminished stellar wind mass loss rates result in a greater retention of initial angular momentum in stars. The enhanced retention of rotational energy leads to increased internal mixing, providing additional hydrogen fuel to the stellar core. As a consequence, when these stars reach the end of their main sequence lifetimes, they develop larger helium cores. This rejuvenation process enabled by rotation plays a crucial role in the modeling of SPS. The metallicity of stars also has a profound effect on their colors and magnitudes. This is primarily due to the influence of metals on the overall energy distribution through \modified{radiative opacity}. Furthermore, metallicity affects the structure and evolution of stars by increasing their opacity, resulting in higher mass loss rates, shorter lifetimes, and different evolutionary tracks compared to stars with lower metallicities.

\begin{figure*}[ht!]
\centering
\includegraphics{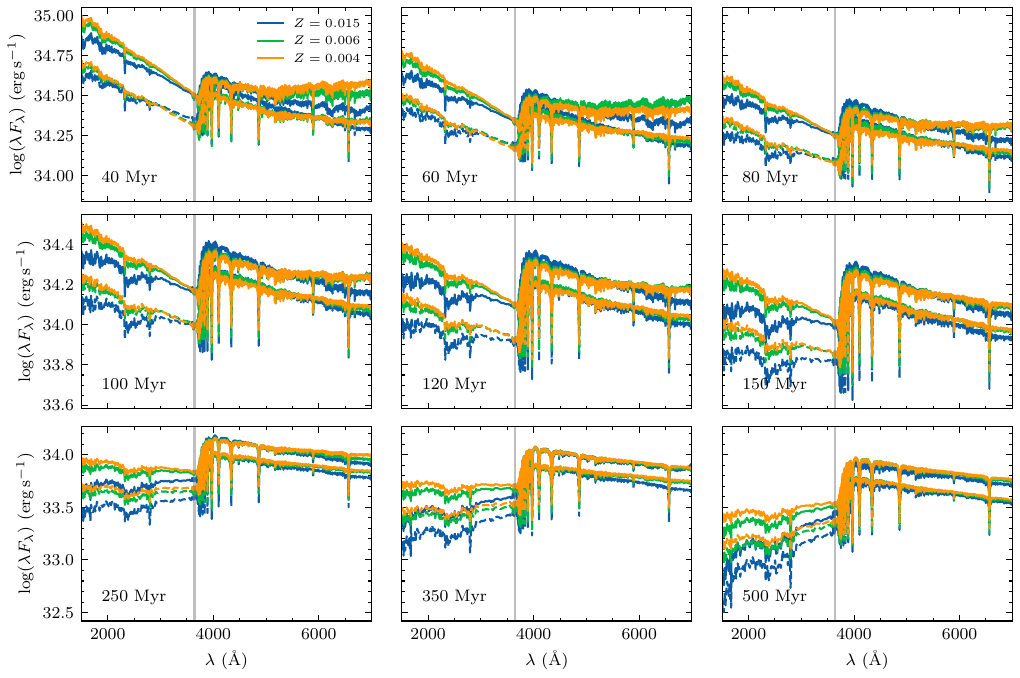}
\caption{Same as Figure~\ref{fig:geneva_spec_rot}, but for PARSEC models with different metallicities ($Z = 0.015$: blue, $Z = 0.006$: green, $Z = 0.004$: orange). The solid lines represent rapid-rotating models ($\omega=0.90$), while the dashed lines depict non-rotating models ($\omega=0.00$). \label{fig:parsec_spec_rot_metal}}
\end{figure*}

In Fig.~\ref{fig:parsec_spec_rot_metal}, we present the time evolution of SED predictions for PARSEC models with different metallicities. The solid lines represent rapid-rotating models ($\omega=0.90$), while the dashed lines represent non-rotating models ($\omega=0.00$). At a fixed age, increasing metallicity has the primary effect of reddening the spectrum. This effect is most apparent in the UV regime, where the total flux decreases as the metallicities increase from $Z=0.004$ to $Z=0.015$. The reason for this behavior is that metal-poor stars evolve at higher effective temperatures and higher luminosities, assuming a fixed initial stellar mass \citep{1992A&AS...96..269S, 2000A&AS..141..371G}. However, the trend becomes less pronounced in the optical band or even reverses at the very early phase of evolution (around $t \sim \unit[100]{Myr}$, as shown in the top row of Fig.~\ref{fig:parsec_spec_rot_metal}). This observation has already been noted by \citet{2003MNRAS.344.1000B}, where they found that the color ($B-V$) of SPS starts to increase with metallicity only after an age of $t \sim \unit[100]{Myr}$ (see their Fig.~1).

\begin{figure*}[ht!]
\centering
\includegraphics{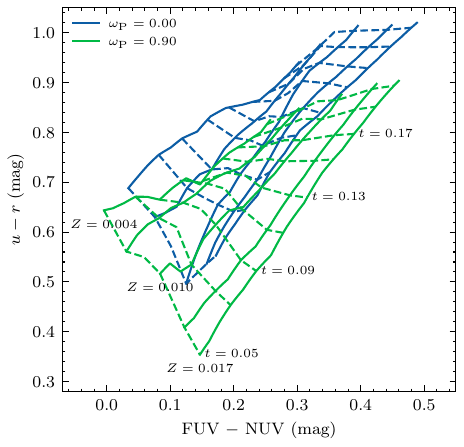}
\includegraphics{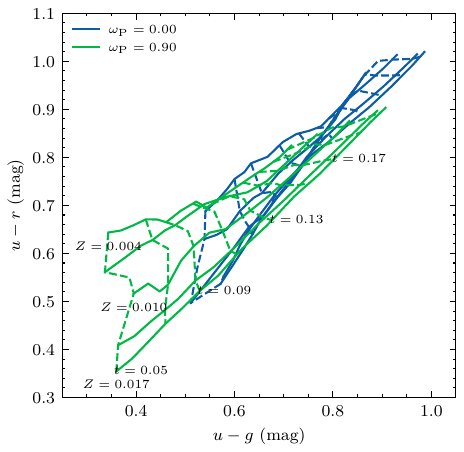}\caption{$u-r$ versus $\mathrm{FUV} - \mathrm{NUV}$ (left) and $u-g$ (right) color grids for non-rotating ($\omega=0.00$) and rapid-rotating ($\omega=0.90$) models. The continuous lines represent iso-metallicity loci for $Z=0.004$, 0.006, 0.01, 0.014, and 0.017, as labeled at the bottom. The dashed lines depict iso-age loci ranging from $t=\unit[0.05]{Gyr}$ to \unit[0.21]{Gyr}, in steps of \unit[0.02]{Gyr}, as labeled on the right. \label{fig:phot_metal}}
\end{figure*}

Such behavior is more clearly illustrated in the left panel of Fig.~\ref{fig:phot_metal}, which presents the grids of model colors, $u-g$ versus $\mathrm{FUV} - \mathrm{NUV}$, for models of various ages and metallicities. For SPS with ages younger than $t\sim \unit[0.15]{Gyr}$, the dashed lines (representing mono-age) exhibit negative slopes, indicating that the $u-r$ colors become redder while the $\mathrm{FUV}-\mathrm{NUV}$ colors become bluer as the metallicity decreases. We set the upper age limit of the grid to be \unit[0.21]{Gyr} since older grids ($t > \unit[0.4]{Gyr}$) would show a strong degeneracy between age and metallicities. In the right panel of Fig.~\ref{fig:phot_metal}, we present a similar diagram but for the color $u-r$ versus $u-g$. As the wavelength domain shifts to the optical range, the degeneracy between age and metallicity becomes more pronounced, and this is further complicated by the involvement of rotation.

Despite the expected decrease in mass-loss rates with rotation as metallicities decrease, the enhancement of the SED caused by rapid rotation remains significant. This is evident in both the spectroscopic evolution, as illustrated in Fig.~\ref{fig:parsec_spec_rot} and \ref{fig:parsec_spec_rot_metal}, and the photometric evolution, as shown in Fig.~\ref{fig:phot_metal}. 

\section{Comparison with observations of star clusters}
\label{sec:cluster}
One crucial step in establishing the reliability of our SPS models is to assess their accuracy in reproducing observed properties, particularly those of star clusters. As mentioned in Section~\ref{sec:intro}, this study was inspired by the discovery of universal rotational effects in both cluster and field stars. In this section, we will primarily focus on the spectroscopic perspective rather than the photometric perspective. While the latter view is also essential, extensive research has already demonstrated the necessity of including rotation to reproduce the CMDs of star clusters in the Milky Way \citep{2018ApJ...869..139C} and Magellanic Clouds \citep{2023A&A...672A.161M}. \modified{Building upon this understanding, several studies, including \citet{2013ApJ...776..112Y, 2014A&A...566A..21G, 2022A&A...665A.126N}, have validated the accuracy of rotational models based on CMDs of clusters. However, it's crucial to recognize that this perspective might be too simplistic for certain stellar environments. As elucidated by \citet{2017ApJ...846...22G, 2019A&A...631A.128C}, for young massive star clusters, accounting for rotation alone may not suffice. These studies underscore the necessity of considering multiple populations with a range of ages, in conjunction with different rotation rates, to comprehensively reproduce the CMDs of these clusters.} Here, we aim to determine whether the proposed SPS model can successfully reproduce the integrated spectra of star clusters.

We use the integrated light spectra from the WAGGS project \citep[WiFeS Atlas of Galactic Globular cluster Spectra]{2017MNRAS.468.3828U}. The WAGGS project observed 64 Milky Way globular clusters and 22 globular clusters hosted by the Milky Way's low-mass satellite galaxies using the WiFeS integral field spectrograph on the Australian National University \unit[2.3]{m} telescope. They provided spectra with a wide wavelength coverage ($\unit[3300-9050]{\AA}$), a moderate spectral resolution ($R = 6800$), and a signal-to-noise ratio of 80 (per \AA) in the NaD region. The WAGGS spectra were convolved with an additional instrumental broadening profile to account for the lower resolution power (\unit[2.5]{\AA} from $\unit[3525-7500]{\AA}$) of the MILES empirical library \citep{2006MNRAS.371..703S, 2011A&A...532A..95F} used in SPS. Normalization was applied to both the WAGGS and SPS spectra through iterative fitting with a spline function, rejecting pixels deviating by more than $\unit[3]{\sigma}$ from the median values in \unit[100]{\AA} windows.

We derived the line-of-sight velocities ($v_\mathrm{los}$) using the cross-correlation function method and estimated the best-fit cluster parameters (cluster age, $t$, and metallicity, $Z$) by minimizing the $\chi^2$ of the normalized spectra. The fitting was performed on a grid of metallicities provided by the models. Two sets of wavelength ranges are used for the spectrum fitting. The first set covered the full wavelength range from \unit[3700]{\AA} to \unit[7350]{\AA}, while the second set covered a narrower range from \unit[3700]{\AA} to \unit[5000]{\AA}. \citet{2020MNRAS.498.2814A} found that the narrower range is better for age determination because including more spectral range on the long-wavelength end of the optical range dilutes the power of the \unit[4000]{\AA} Balmer jump. For the first set, we also masked the wavelength range around H$\alpha$ (from \unit[6540]{\AA} to \unit[6593]{\AA}) to account for emission line features in young clusters (e.g., NGC 1850), which are not included in the MILES stellar library. These H$\alpha$ emissions are connected with Be stars, in which the emission arises from a Keplerian decretion disk surrounding a rapidly rotating star \citep{2012MNRAS.425..355R, 2013A&ARv..21...69R}. Moreover, these emissions at H$\alpha$ are confirmed to originate from fast-rotating stars populating the red part of the MS or eMSTO, as verified through narrow-band imaging \citep[e.g.,]{2018MNRAS.477.2640M} and single-star spectroscopy \citep{2017ApJ...846L...1D}.

\begin{figure*}[ht!]
\centering
\includegraphics{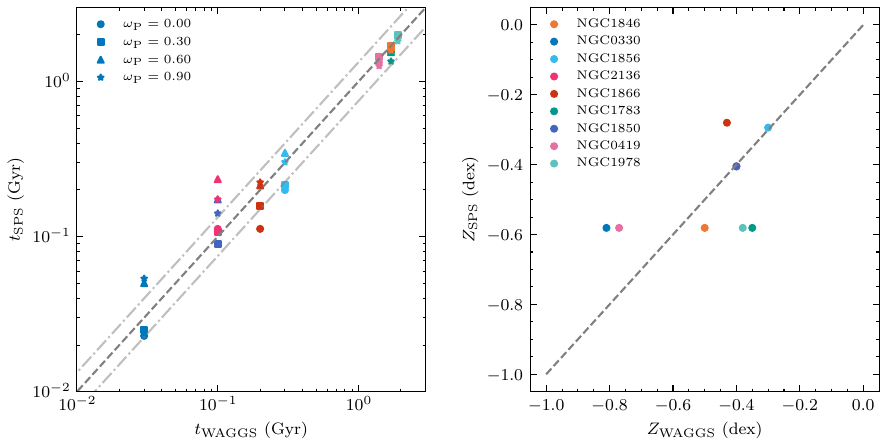}
\caption{Comparison of cluster age, $t$ (left), and metallicity, $Z$ (right), derived from SPS models with rotation and literature results of clusters in WAGGS \citep{2017MNRAS.468.3828U}. The spectrum fitting was based on the wavelength range $3700 \leqslant \lambda/\unit{\AA} \leqslant 5000$. For this comparison, we selected clusters of interest from the Large and Small Magellanic Clouds. The horizontal axis presents the values collected from \citet{2007AJ....134.1813M, 2008AJ....136..375M, 2009AJ....138.1403G, 2011MNRAS.413..837M, 2012ApJ...746L..19M, 2013MNRAS.431L.122B, 2014ApJ...797...35G, 2015A&A...575A..62N}. Different colors represent different clusters, and markers represent SPS models with various rotation rates. The median values of the best-fit metallicity are shown for each cluster. Both the cluster age and metallicity exhibit a one-to-one correlation (grey dashed line) with the literature results. An age uncertainty of \unit[0.1]{dex} (25\%) is shown in the left panel as dash-dotted lines. NGC2136 shares the same location as NGC 1850. \label{fig:compare_cluster}}
\end{figure*}

The derived $v_\mathrm{los}$ are consistent with the tests in the literature \citep[e.g.,][]{1992AJ....103..447S, 2006AJ....132.1630G, 2019MNRAS.484.2832V}. In Fig.~\ref{fig:compare_cluster}, we compare the cluster age, $t$, and metallicity, $Z$, estimated from the SPS models with rotation to the literature results of clusters in WAGGS \citep{2017MNRAS.468.3828U}. Two sets of analyses based on different wavelength ranges yield similar results, and only the case of $3700 \leqslant \lambda/\unit{\AA} \leqslant 5000$ is presented here. To investigate the possible influence of stellar rotation, we focus on clusters younger than $\sim\unit[3]{Gyr}$, all of which are located in the Milky Way satellites, the Large and Small Magellanic Clouds. In both panels, colors indicate different clusters, and markers represent various models with rotation rates ($\omega_\mathrm{P}=0.00, 0.30, 0.60, 0.90$) in the left panel. For metallicity, the median values of the best-fit values are shown, considering a coarse grid with $Z = -0.58, -0.40, -0.28, -0.14, 0.04$, and $0.05$. Due to the metal-poor end of the grid in the metallicity range of PARSEC, the metallicities of NGC 419 and NGC 330 (pink and dark blue dots in Fig.~\ref{fig:compare_cluster}) are biased toward more metal-rich values compared to the literature results ($Z_\mathrm{WAGGS} = -0.77$ for NGC 419 and $Z_\mathrm{WAGGS} = -0.81$ for NGC 330). However, despite this deviation, the majority of the metallicity estimates derived from SPS modeling fall within the range of approximately $\unit[-0.3]{dex}$ to $\unit[-0.6]{dex}$, which is consistent with results obtained from single-star high-resolution spectra.

Regarding cluster age, different stellar rotation models also yield similar estimates for the clusters. This is because the wavelength coverage of the WAGGS spectra only includes the optical range, where the differences between various degrees of rotational effects are not as significant compared to the UV range (see Fig.~\ref{fig:parsec_spec_rot_metal}) and the divergence between metallicity and age becomes more pronounced (see the right panel of Fig.~\ref{fig:phot_metal}). However, there are indeed some differences in the predicted cluster ages among various models, and this could be a probe to test the robustness of rotational models.

SPS modeling with small rotation rates (represented by circles for $\omega_\mathrm{P} = 0.00$ and squares for $\omega_\mathrm{P} = 0.30$ in the left panel of Fig.~\ref{fig:compare_cluster}) generally yields younger ages compared to models with larger rotation rates. However, this difference diminishes as the clusters become older than $\unit[1]{Gyr}$. Nevertheless, the accuracy of age determination for these clusters from photometric observations limits the effectiveness of this test. While most of these clusters have high-quality observations from the \textit{Hubble Space Telescope} and their ages are estimated through the isochrone-fitting technique, the determination of their intrinsic ages is hindered by the complexities of the stars around their main-sequence turnoff regions (as mentioned in Section~\ref{sec:intro}). Additionally, full-spectrum fitting results in larger age uncertainties in the range of $8.2 \leqslant \log (t/\unit{yr}) \leqslant 8.8$ \citep{2020MNRAS.498.2814A} since the shape of the color-magnitude diagram does not significantly change with age in these intervals. The typical precision of \unit[0.1]{dex} (25\%) estimated by \citet{2020MNRAS.498.2814A} for $\mathrm{S/N} \geqslant 50$ in the age range of $7.0 \leqslant \log (t/\unit{yr}) \leqslant 9.5$ is smaller than the variation seen here, which represents the intrinsic systematic differences caused by rotation. This finding is consistent with the results discussed in Section~\ref{sec:rotation}, particularly in the age range where the rotational effects are prominent.

\section{Discussion}
\label{sec:discussion}
We have illustrated the effect of stellar rotation on SPS models and demonstrated its effectiveness through comparisons with observed integrated spectra of star clusters with varying ages and metallicities. We now shift our focus to the predictions of our models for young galaxies ($t < \unit[500]{Myr}$) observed in the high-redshift universe and discuss the inferred properties of these galaxies from JWST multi-band observations. In Section~\ref{sec:galaxy_model}, we describe how we construct the galaxy model using the \texttt{Prospector} code and present the predictions for stellar rotation based on mock data. Then, in Section~\ref{sec:galaxy_obs}, we compare our model with observed high-redshift galaxy photometry from JWST and discuss the potential biases in current interpretations.

\subsection{Building galaxy models}
\label{sec:galaxy_model}
The galaxy stellar population models are constructed using the \texttt{Prospector} code \citep{2021ApJS..254...22J}, a fully Bayesian inference tool for deriving stellar population properties from photometric and spectroscopic data. Recent works \citep[e.g.,][]{2022ApJ...927..170T, 2023NatAs...7..611R} have extended its usage from mainly low-redshift galaxies to higher redshifts ($z\sim 10$). We utilize the PARSEC models, together with the \texttt{FSPS} package (see Section~\ref{sec:sps}). Throughout this work, we assume a \citet{2003PASP..115..763C} initial mass function.

We model the SED with a 13-parameter model that includes stellar mass (uniform distribution between $7<\log(M_\star/M_\odot)<12$) and stellar metallicity (a clipped normal prior in $\log(Z/Z_\odot)$ with mean $\mu=-1.5$ and standard deviation $\sigma = 0.5$, with a minimum and maximum of -2.0 and 0.2, respectively). Furthermore, we adopt a flexible SFH prescription with 8 time bins. The SFH of the galaxies is assumed to start at redshift $z=20$, and the time bins are log-spaced out to $z=20$. This SFH prescription follows a constant SFR prior, where the logarithm of the ratio of the SFRs of two adjacent time bins follows a Student's t-distribution with a mean $\mu=0$ and a standard deviation of $\sigma=1$. The most recent two age bins are fixed at lookback times of $0-5$ and $\unit[5-10]{Myr}$ in the SFH to capture recent bursts that may be powering extreme nebular emission \citep{2023MNRAS.519..157W}.

\modified{Following the approach in \citet{2004MNRAS.351.1151B, 2009A&A...507.1793N, 2011MNRAS.417.1760W, 2022ApJ...927..170T}}, we adopt a two-component dust attenuation model. The first component is a diffuse component that has a variable attenuation curve and attenuates all stellar and nebular emissions from the galaxy. The optical depth at \unit[5500]{\AA}, denoted as $\hat{\tau}_{\text{dust, 2}}$, is assigned a flat prior between 0.0 and 1.0 for this component. We adopt the prescription from \citet{2009A&A...507.1793N} and a \citet{2013ApJ...775L..16K} attenuation curve for this component. The curve's slope ($n$), known as the dust index, is a free parameter modeled as an offset from the slope of the UV attenuation curve by \citet{2000ApJ...533..682C}, with a flat prior between $-1.0$ and $0.4$. The second component is a birth-cloud attenuation model that operates in the last \unit[10]{Myr}. It is characterized by a number ratio between the birth cloud and diffuse dust attenuations. The combination of these two components allows for a more comprehensive modeling of the dust properties within the galaxy.

The nebular emission lines and continuum are self-consistently modeled \citep{2017ApJ...840...44B}. The gas-phase metallicity ($Z_\text{gas}$) is tied to the stellar metallicity, and a fixed ionization parameter of $\log U = -2.25$ is adopted. For spectra bluer than $\lambda < \unit[1216]{\AA}$ in the rest frame, we include a $z$-dependent intergalactic medium (IGM) attenuation following \citet{1995ApJ...441...18M}. A free parameter, denoted as $f_\text{IGM}$, scales the total IGM opacity to account for line-of-sight variations in the total opacity. The prior for $f_\text{IGM}$ is a clipped Gaussian distribution centered on 1, with a dispersion of 0.3 and clipped at 0 and 2. This parameter provides flexibility in capturing the variations in the IGM opacity along different lines of sight.

We simulate mock galaxies with redshifts ranging from $z=8$ to 12 and total mass ranging from $\log(M/M_\odot)=8.6$ to 9.0. The remaining input parameters are chosen from the mean values of an initial guess, except for $A_V$, which was set to 0.2. Each set of configurations is simulated 50 times with a minor scatter in the input parameters to ensure reasonable sampling. To mimic real observations (see Section~\ref{sec:galaxy_obs}), a static error of \unit[3]{nJy} is adopted for all wide filters, including those with flux densities weaker than this value. The posteriors are sampled using the dynamic nested sampling code \texttt{dynesty} \citep{2020MNRAS.493.3132S}.

\begin{figure}[ht!]
\centering
\includegraphics[width=90mm]{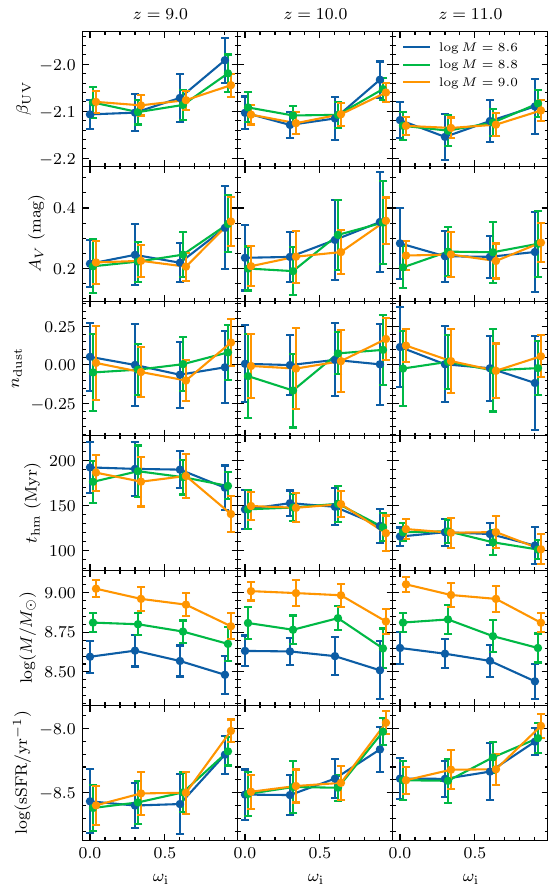}
\caption{Dependence of the derived physical quantities on the initial rotation rates, showing UV slope ($\beta$), dust attenuation in the $V$ band ($A_V$), dust index ($n$), half-mass time (the lookback time at which 50\% of the stellar mass formed), stellar mass, and specific star formation rate. The mock galaxies are generated with an SPS model of a non-rotating model ($\omega_i = 0.0$), and the physical quantities of these mock galaxies are then estimated with models of different rotation rates ($\omega_i = 0.0$, 0.3, 0.6, and 0.9). For each quantity, the results are grouped into three redshifts from $z=9.0$ to $11.0$. The colored curves represent the results binned by $\log(M/M_\odot)$, with legends shown in the top-right subplot. A minor shift is applied to $\omega_\text{i}$ for better visualization. \label{fig:mock_trend}} 
\end{figure}

To explore the influence of different SPS models, we generate mock galaxies using non-rotating SPS models. The posteriors and derived physical quantities are then estimated with different rotation rates ($\omega_i = 0.0$, 0.3, 0.6, and 0.9). We investigate six properties: UV spectral slope ($\beta_\text{UV}$), dust attenuation ($A_V$), dust index ($n$), stellar age ($t_\text{hm}$), stellar mass ($\log(M/M_\odot)$), and specific star formation rate (sSFR). The UV spectral slope, $\beta$, defined as the power-law index $f_\lambda \propto \lambda^{\beta}$ \citep{1994ApJ...429..582C}, is estimated from the best-fit \texttt{Prospector} spectra following the method of \citet{2022ApJ...927..170T}. The stellar age is inferred from the lookback time at which 50\% of the stellar mass formed. These properties are separated into three redshift bins (columns) from $z=9.0$ to 11.0 and shown as a function of the rotation rates of the SPS model in Fig.~\ref{fig:mock_trend}. In each subplot, the trend was binned by stellar mass ($\log(M/M_\odot) = 8.6$, 8.8, and 9.0). A minor shift is applied to $\omega_\text{i}$ to avoid overlapping. The error bars indicate the $\unit[1]{\sigma}$ scatter from bootstrapping. It should be noted that the absolute values of the test are not our focus since they are mainly determined by the reference model we choose. Instead, our emphasis is on the variation of these properties and their dependence on rotation.

The first three rows focus on the dust properties. $\beta$ is a direct measure of the color of the emerging light from the young, massive stars in these early galaxies. While stellar population age and stellar metallicity can also affect the rest-frame UV color, the observed slope $\beta$ is much more sensitive to dust attenuation \citep{2022ApJ...927..170T}, and it is generally interpreted as a proxy for dust attenuation. The $\beta$ parameter is also correlated with $A_V$ (the level of dust attenuation at \unit[5500]{\AA}) and $n_\text{dust}$ (linked to the strength of the UV bump). We observe a significant increase in the recovered $\beta_\text{UV}$ as rotation rates increase, amounting to up to \unit[0.1]{dex}. These trends remain consistent across the range of redshift and stellar mass studied, although the scale of the effect is relatively smaller for $z=11.0$. Regarding $A_V$ and $n_\text{dust}$, it is challenging to discern the effects, although higher values of $A_V$ are noticeable for $\omega_\text{i}\geqslant 0.6$. This is because SPS models with rapid rotation produce steeper UV slopes compared to non-rotating models. Consequently, to account for the excess of UV photons, a higher degree of dust attenuation is preferred, resulting in larger $A_V$ and $\beta_\text{UV}$ values. 

\begin{figure}[ht!]
\centering
\includegraphics{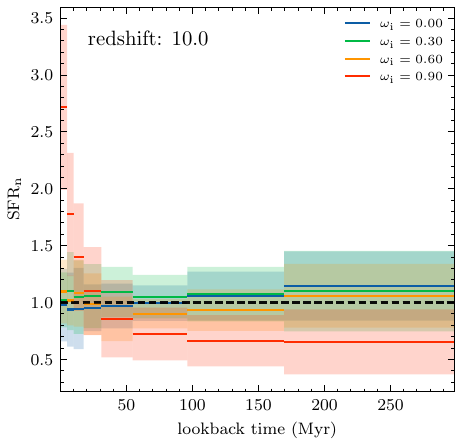}
\caption{Recovered normalized SFR ($\text{SFR}_\text{n}$) from SPS models with different rotation rates, as shown in Figure~\ref{fig:mock_trend}. The input data were generated for a mock galaxy with $\log(M/M_\odot)=9.0$ at a redshift of $z=10.0$ with a constant SFR. Different colors represent the recovered SFR from different SPS models, normalized by the input SFH. The time bins are log-spaced out to $z=20$, assuming that the SFH of the galaxies starts at that redshift. The most recent two age bins are fixed at lookback times of $0-5$ and $\unit[5-10]{Myr}$. \label{fig:mock_sfr}} 
\end{figure}

The last three rows of the figure present the variation of properties linked to star formation history. SPS models with rotation rates $\omega_\text{i}\leqslant 0.6$ predict similar results for half-mass time ($t_\text{hm}$) and specific star formation rate (sSFR). However, as the rotation rates of the stellar population increase to $\omega_\text{i}=0.9$, there is a significant decrease in $t_\text{hm}$ and stellar mass, accompanied by an increase in sSFR. The reason is that the production of more rest-frame UV fluxes than rest-frame optical fluxes, as seen in Fig.~\ref{fig:phot_rot} and \ref{fig:parsec_spec_rot_metal}, biases the stellar mass obtained from UV-derived SFR estimates towards lower values in the case of rapid rotation. This suggests that, with an SPS model with rapid rotation ($\omega_\text{i}=0.9$), the stellar masses could be around $20\%-40\%$ smaller than those inferred with a non-rotating model, and a recent starburst activity could be favored, characterized by a stronger sSFR of up to four times. This trend in $\log(\text{sSFR})$ persists even when the SFR is averaged over periods of $\unit[30]{Myr}$ ($\text{SFR}_\text{30}$) or $\unit[50]{Myr}$. The opposite pattern in $t_\text{hm}$ and $\log(\text{sSFR})$ is more clearly demonstrated in Fig.~\ref{fig:mock_sfr}, where the recovered star formation histories (SFHs) are shown for different models and normalized by the input SFH at a redshift of $10.0$ with $\log(M/M_\odot) = 9.0$. The model with $\omega = 0.90$ exhibits a pronounced recent starburst within the last $\unit[30-50]{Myr}$, while for models with moderate rotation rates, this bias is much less significant. This would explain the sharp increase in sSFR and the decrease in half-mass time observed in Fig.~\ref{fig:mock_trend}. \modified{Based on the posterior distribution of dust attenuation versus half-mass age, we confirmed that there is an anti-correlation between older ages and more UV attenuation, indicating it is challenging to break the dust-age degeneracy with our current observational data \citep[e.g.,][]{2001ApJ...559..620P, 2022ApJ...927..170T}}

\subsection{Interpretation of high-redshift galaxy observations}
\label{sec:galaxy_obs}
In this section, we exemplify how our model can be used to interpret observed galaxy spectra. We consider an observational sample of five spectroscopically confirmed high-redshift ($z > 10$) galaxies observed by JWST. This sample was obtained through the JADES survey \citep{2020IAUS..352..342B}, which successfully measured a selection of galaxies with confirmed redshifts at $z > 10$. Physical properties of four of these galaxies, namely GS-z10-0, GS-z11-0, GS-z12-0, and GS-z13-0, were determined based on their mean spectroscopic redshifts of $z = 10.38$, $z = 11.58$, $z = 12.63$, and $z = 13.20$, respectively, as described in the research works by \citet{2022ARA&A..60..121R, 2023NatAs...7..611R, 2023NatAs...7..622C, 2023A&A...677A..88B}. Additionally, the spectroscopic follow-up \citep{2023Natur.622..707A} of high-redshift candidates reported by the Cosmic Evolution Early Release Science project \citep[CEERS,][]{2023ApJ...946L..13F} confirmed the redshift of CEERS2-5429 (published as Maisie's Galaxy by \citet{2022ApJ...928...52F}) at $z = 11.44$. NIRSpec follow-up observations were used to confirm the redshift of the NIRCam-identified sources in both surveys. The sources for this analysis have at least eight NIRCam photometric bands, mostly wide-band filters spanning approximately $\unit[1-5]{\mu m}$. In the case of GS-z11-0, additional F182M, F210M, F430M, F460M, and F480M images are available as part of the JWST Extragalactic Medium-band Survey (JEMS) by \citet{2023ApJS..268...64W}, which provides further characterization of this galaxy.

\begin{figure*}[ht!]
\centering
\includegraphics{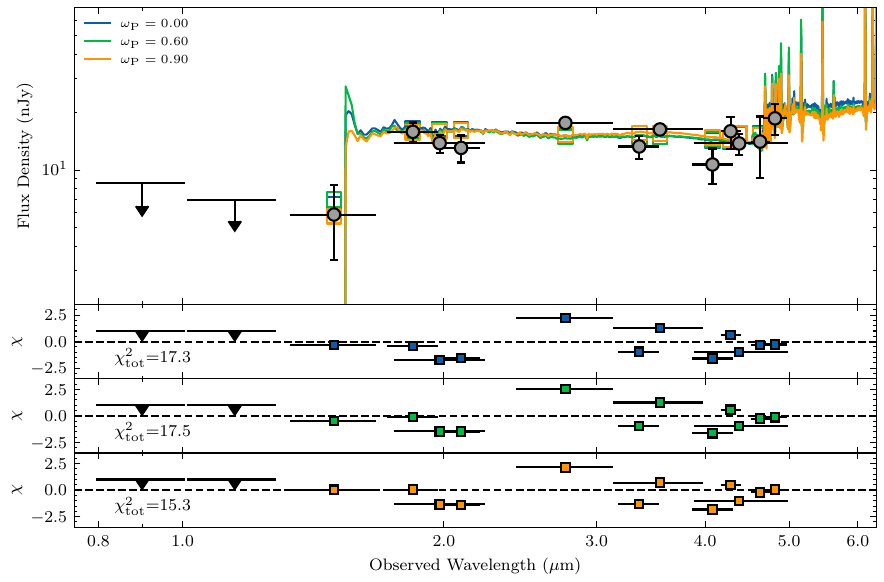}
\caption{SED modeling of galaxy GS-z11-0 ($z=11.58$) with PARSEC models incorporating different rotation rates. The observational data used in the SED fitting is shown in the top panels. The detected filters are represented by gray circles with horizontal error bars indicating the approximate wavelength range of each filter. For passbands below the detection limit, their upper limits are indicated by bars with an arrow pointing down at $5\sigma$ significance. The blue, green, and orange lines represent the best-fitting SEDs predicted from PARSEC models with rotation rates of $\omega_\text{P,i} = 0.0$, 0.6, and 0.9, respectively. The boxes mark the posterior fluxes in the different filters. The bottom panels show the $\chi$ values for each filter, using the same color scheme as the top panel. The total $\chi^2$ value is given at the bottom left of the bottom panels.  \label{fig:GSz1158_sed}}
\end{figure*}

In Fig.~\ref{fig:GSz1158_sed}, we present examples of the best-fit spectral energy distributions (SEDs) of GS-z11-0 using non-rotating models ($\omega_\text{P,i} = 0.00$, blue) compared to those with moderate rotation ($\omega_\text{P,i} = 0.30$, green) and rapid rotation ($\omega_\text{P,i} = 0.90$, yellow). The residuals, defined as $\chi=(F_\text{model}-F_\text{obs})/\sigma_\text{obs}$, are centered around $0$, indicating that the data are well-fit by the model. Moreover, the residuals among the different models exhibit nearly identical patterns. This suggests that even for galaxies with comprehensive passband coverage, the observations cannot differentiate between models with different rotation rates.

\begin{figure*}[ht!]
\centering
\includegraphics{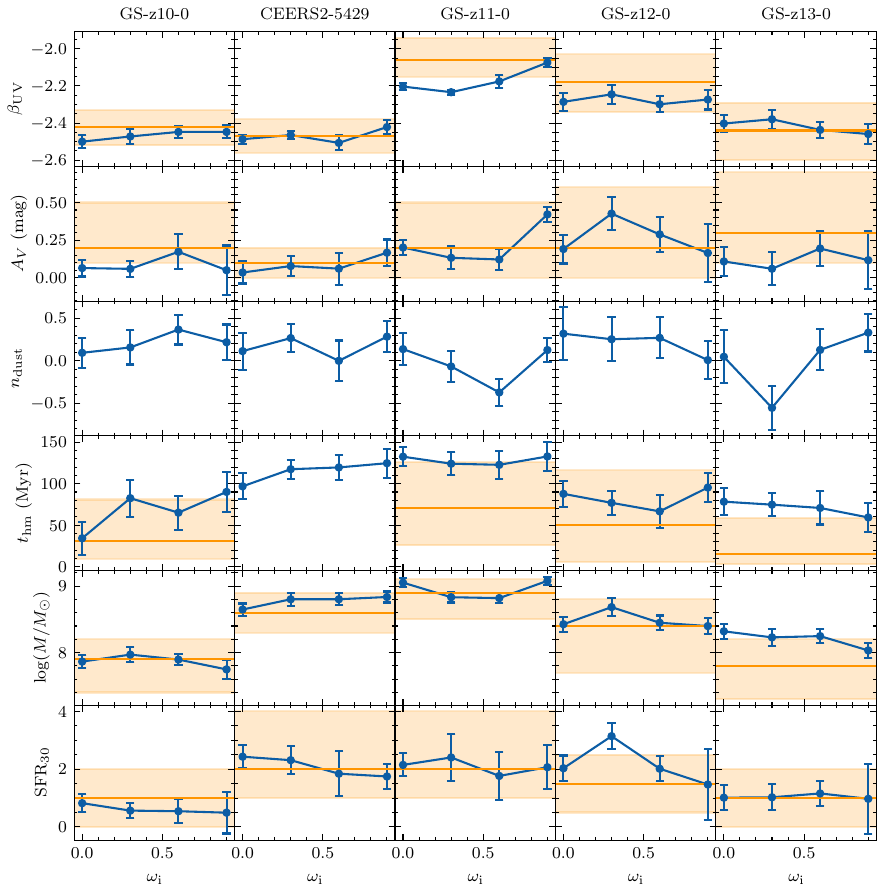}
\caption{Derived physical quantities of five spectroscopically-confirmed high-redshift galaxies from SPS models with different rotation rates. The galaxies are ordered by their spectroscopic redshift ($z_\text{spec}$) from left to right. The error of the derived quantities was estimated via bootstrapping, as described in Figure~\ref{fig:mock_trend}. The orange horizontal lines and the shadowed region around them represent the best-fit results and $\unit[1]{\sigma}$ uncertainties obtained from \citet{2023NatAs...7..611R, 2023Natur.622..707A}, which were fitted using \texttt{Prospector} with the MIST stellar evolution models \citep{2016ApJ...823..102C}. \label{fig:obs_trend}}
\end{figure*}

The results of fitting with different rotation models are summarized in Fig.~\ref{fig:obs_trend}, where the same properties as in Fig.~\ref{fig:mock_trend} are investigated, except $\text{SFR}_{30}$. We compare the inferred values from our model with those collected from \citet{2023NatAs...7..611R}, where non-rotating MIST stellar evolutionary tracks and isochrones \citep{2016ApJ...823..102C} were employed. The best-fit results and their corresponding errors are shown as orange horizontal lines and the orange shadowed regions. It is important to note that the physical quantities are estimated using different rotational models in this figure. This differs from Fig.~\ref{fig:mock_trend}, where the mock galaxies were generated using different models but fitted using the same non-rotating model.

We find that the derived physical properties are consistent with the results in the literature within approximately $\unit[1]{\sigma}$. The uncertainties of our estimates are smaller than those derived from the nested sampling approach because the bootstrapping process does not consider the variation in redshift. For the SFH-related parameters (the bottom three rows), we observe no significant variation as a function of rotation rates (except for $t_\text{hm}$ of GS-z10-0). The largest deviation appears for GS-z13-0, which has the highest redshift. This discrepancy might be due to the improper handling of massive stars in our models. As the PARSEC models with rotation only reach a stellar mass of $\approx\unit[15]{M_\odot}$, we merged the non-rotating tracks with the rotation models for the massive star regime to ensure reasonable estimates for stellar mass and SFH. However, this unavoidably introduces a bias for very young ($t<\unit[10]{Myr}$) stellar populations, as observed in Fig.~\ref{fig:specbymass}. This influence would quickly diminish as the ages of the galaxies increase, but it might remain noticeable for galaxies with redshifts $z>13.0$, such as GS-z13-0. For the dust-related parameters, all models predict a low level of dust attenuation in these galaxies. The variation in $A_V$ and $n_\text{dust}$ is minor compared to their uncertainties. However, there is a tendency in $\beta_\text{UV}$ where the UV spectral slope might increase as the SPS rotation rates increase, although this effect is not significant compared to the associated uncertainty in real data (as indicated by the orange shadowed region in Fig.~\ref{fig:obs_trend}).

\begin{figure}[ht!]
\centering
\includegraphics[width=90mm]{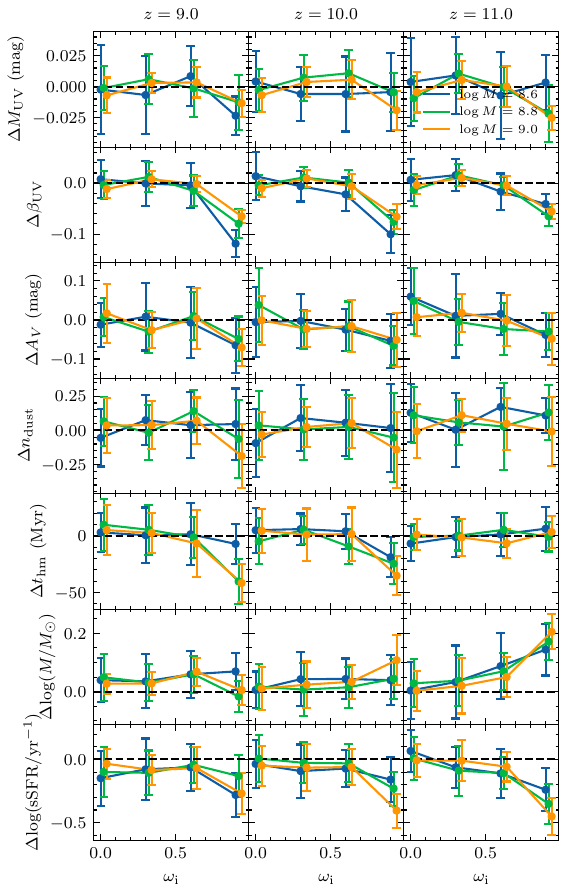}
\caption{Similar to Figure~\ref{fig:mock_trend}, but showing the differences ($\Delta X \equiv X - X_\text{input}$) between the derived and input physical parameters. The mock galaxies were generated based on SPS models with different rotation rates but fitted using a non-rotating SPS model. $\Delta = 0$ is represented by the horizontal dashed lines.
\label{fig:mock_nonrot_delta_trend}} 
\end{figure}

Despite the potential bias introduced by interpreting the data with different SPS models, unfortunately, we cannot definitively identify which SPS model is the correct prescription based solely on the observed data. However, although the models' variation for individual galaxies might be smaller than the associated uncertainty, it would have a systematic bias from the true values and it is still meaningful to explore the possible deviations that may arise when modeling with a wrong SPS model. In this context, we specifically refer to the widely used non-rotating models, which serve as the baseline for SPS modeling. Hence, we conduct a similar simulation as mentioned in Section~\ref{sec:galaxy_model}, but this time we generate multiple mock galaxies using SPS models with different rotation rates and fit these mock samples with a non-rotating SPS model. The derived physical parameters are then compared with the input mock galaxies, and their differences ($\Delta X \equiv X - X_\text{input}$) are shown in Fig.~\ref{fig:mock_nonrot_delta_trend}.

As already shown in Fig.~\ref{fig:mock_trend} and Fig.~\ref{fig:obs_trend}, a non-rotating SPS model can accurately reproduce the physical properties of the input mock galaxy up to a rotation rate of $\omega_\text{i} = 0.6$. However, when the rotation rates increase to $\omega_\text{i} = 0.9$, the non-rotating model becomes unable to accurately describe the system, particularly in terms of $M_\text{UV}$, $\beta_\text{UV}$, $t_\text{hm}$, and sSFR. The UV spectral slope ($\beta_\text{UV}$) could be underestimated by approximately \unit[0.1]{dex}, and the stellar age ($t_\text{hm}$) is underestimated by up to \unit[40]{Myr}. It is important to note that this bias in stellar age is only significant for the mock galaxies at redshifts $z=9.0$ and $z=10.0$, while there is no significant bias at $z=11.0$. This disparity may be coincidental, as an underestimation of recent SFR may be counterbalanced by an overestimation of past SFR, resulting in a stellar age estimate that is less sensitive to the change in rotation rates. There is no clear trend observed in $\Delta n_\text{dust}$, likely due to the relatively large uncertainties associated with these measurements. We also see an increasing bias in stellar mass and sSFR as the redshift goes higher than 10.

A more realistic scenario arises when the stellar population is a mixture of populations with various rotation rates. This has been confirmed both in clusters/star associations and in nearby field stars. While the effect of rotation has been proven necessary to reproduce the light from clusters \modified{\citep[but see also][for evidence of multiple populations in NGC 1866]{2019A&A...631A.128C}}, it is widely acknowledged that the rotation rates/velocities of the cluster members are not identical. For example, \citet{2019ApJ...883..182S} found that stars belonging to the blue and red main sequences in the open cluster NGC 2287 exhibit distinct rotational velocities. Similarly, \citet{2020MNRAS.492.2177K} revealed a distribution of rotation rates in NGC 1846, supporting a bimodal distribution, with fast rotators centered around $v\sin i = \unit[140]{km\,s^{-1}}$ and slow rotators centered around $v\sin i = \unit[60]{km\,s^{-1}}$. For clusters with ages around \unit[1]{Gyr}, \citet{2021MNRAS.502.4350S} found significant variations in the MSTO both in the peak locations and the overall distributions on the CMD, suggesting a wide distribution of stellar rotation across clusters. Moreover, numerous works have also identified similar bimodal rotational distributions for massive stars in nearby field environments \cite{2010ApJ...722..605H, 2012A&A...537A.120Z, 2021ApJ...921..145S}. If the early galaxies follow a similar distribution as observed in young star clusters and in the Milky Way, it is more likely that their rotation distribution is a combination of various rotation rates. As suggested in Fig.~\ref{fig:mock_nonrot_delta_trend}, the moderate rotation population does not introduce any significant bias when interpreted using non-rotating SPS models. The rotational distribution in these galaxies could be approximated by two populations: one with no rotation and another with rapid rotation. Furthermore, as stars in low-metallicity environments can retain their rotation more effectively, the branch of rapid rotation ($\omega_\text{i} \geqslant 0.8$) could be more pronounced compared to what is commonly observed in local solar-metallicity environments \citep{2021ApJ...921..145S}.

The rest-UV luminosity function (UVLF) serves as a key observational diagnostic of galaxy evolution in the early Universe. \citet{2023ApJ...946L..13F} estimated the UVLF using a sample of 26 candidate galaxies at $z \sim 8.5 - 16.5$ from the CEERS project, finding no significant evolution compared to slightly lower redshifts, which is consistent with the conclusion drawn by \citet{2020MNRAS.493.2059B}. This conclusion was later confirmed by \citet{2023MNRAS.518.6011D}, who reanalyzed the JWST Early Release Observations and Early Release Science imaging surveys, as well as the near-infrared imaging in the COSMOS field provided by Data Release 5 from the UltraVISTA survey \citep{2012A&A...544A.156M}. Based on the lack of bias introduced by the rotating SPS models in Fig.~\ref{fig:mock_nonrot_delta_trend}, the same conclusion would hold even for the most rapidly rotating population.
 
\citet{2023arXiv231106209C} analyzed $\beta_\text{UV}$ for a population of 176 galaxy candidates with $8< z_\text{phot}< 16$, selected using JWST NIRCam imaging and COSMOS/UltraVISTA ground-based near-infrared imaging \citep{2023MNRAS.520...14C}. They found evidence for a relationship between $\beta$ and $M_\text{UV}$, with a slope of $\mathrm{d}\beta/\mathrm{d}M=-0.17\pm 0.03$. It is expected that a similar slope, but possibly with a larger scatter, would be observed when considering the effect of rotation, as we have not found a significant bias in the inferred $M_\text{UV}$. They also discovered a significant UV trend between $\beta$ and redshift, with the mean value evolving from $\langle\beta\rangle=-2.17\pm0.05$ at $z=9.5$ to $\langle\beta\rangle=-2.56\pm0.05$ at $z=11.5$. If the population of these early Universe galaxies could be described by a rapidly rotating population, the actual variation in $\beta$ at different redshifts could be larger, as the bias observed in Fig.~\ref{fig:mock_nonrot_delta_trend} is smaller at higher redshifts. Moreover, the underestimation of $\beta$ at $z \sim 11$ suggests that the galaxies have not yet reached the blue limit of `dust-free' stellar populations \citep{2010Natur.468...49R, 2022MNRAS.517.5104C}, with a margin of approximately $\sim\unit[0.05]{dex}$. Nevertheless, the steeper slope between $\beta$ and $z$ would still indicate a `dust-free' stellar population at a slightly higher redshift.

\modified{Carbonaceous dust fraction is another important factor that may impact the UV spectral index. The production of carbon and the subsequent formation of carbonaceous grains through the asymptotic giant branch (AGB) channel, particularly in the low-metallicity regime characterizing such early galaxies could contribute an excess attenuation known as the UV attenuation `bump' \citep{1965ApJ...142.1683S}. However, such a young population will not carbon-bearing dust grains can form carbon-bearing dust grains. As the minimum initial mass for these AGB progenitors would be over 5 Msun, they experience hot-bottom burning, which destroys the carbon in the envelope. It should be noted that \citet{2023Natur.621..267W} discovered galaxies with large amounts of carbonaceous dust at an age of $\sim\unit[600]{Myr}$, which could not be explained through the standard AGB route. This may suggest another dust production channel in supernovae or Wolf-Rayet stars from high-mass stars. If this also stands for galaxies at a redshift of 10, the extra attenuation at rest-frame \unit[2175]{\AA} could produce a steeper UV slope by \unit[0.06]{dex} \citep[inferred from bump and non-bump sample from][]{2023Natur.621..267W}, which may mask the effects of rotation.}

Regarding the sSFR, \citet{2022ApJ...927..170T} noted that their values for 11 galaxies at $z = 9-11$ are consistent with lower redshift estimates and are slightly below the expected increasing evolution according to theoretical models (see their Figure~10). However, if the bias introduced by rapid rotation is a common feature in these young galaxies, their sSFR could be higher than inferred by approximately $\unit[0.3-0.5]{dex}$, which would align more closely with the predictions from theoretical models \citep{2019MNRAS.488.3143B, 2019MNRAS.490.2855Y}.

\section{Summary}
\label{sec:summary}
In this work, we explore the role of stellar rotation in reducing uncertainties in SPS models. Using a combination of observational data and theoretical models, we investigate the impact of stellar rotation on stellar evolution and spectral properties. Our main conclusions are as follows:
\begin{itemize}
	\item Rapid rotation enhances the SPS flux in both the UV and optical domains. The UV photons show a particularly significant boost as a result of the elongation of the main sequence lifetime for hot stars. This influence is consistent across both PARSEC and GENEVA models and can last up to approximately $\sim\unit[400]{Myr}$, contributing to the stellar population's starburst phase.
	\item At lower metallicities, where stellar winds are less effective in slowing down rotation, SPS models with rotation produce a harder UV slope.
	\item SPS models with rotation can recover the ages of star clusters through integrated light spectra. However, the accuracy of age inferred from isochrone-fitting techniques in the literature hinders the further distinction between different rotation rates.
	\item Using SPS models with different rotation rates ($\omega_\text{i}=0.0$, 0.3, 0.6, 0.9), we make predictions for young galaxies in the high-redshift ($z\sim10$) universe based on JWST multi-band observations. We observe that models with rapid rotation yield a more gradual slope in the UV wavelength range (up to \unit[0.1]{dex}) as well as a higher level of dust attenuation. The stellar mass and half-mass age of these galaxies decreases by $20\%-40\%$ as rotation rates increase to $\omega_\text{i}=0.9$. A constant SFH could be well reproduced by an SPS with rotation rates up to $\omega_\text{i}=0.6$, but a boost of $2-4$ times of the SFH in the last \unit[5]{Myr} is inferred for models with $\omega_\text{i}=0.9$.
	\item For the five high-redshift galaxies with spectroscopic confirmation from JWST, the currently available data cannot differentiate between models with different rotation rates. All models yield dust- and star formation-related physical parameters consistent with literature values within $\unit[1-2]{\sigma}$ uncertainties.
	\item Our analysis demonstrates that for galaxies with rotation rates up to $\omega_i=0.6$, non-rotating models provide accurate estimates of their physical properties. However, beyond this rotation rate, non-rotating models may lead to misinterpretation of certain properties, such as a slight underestimation of \unit[0.1]{dex} in $\beta_\text{UV}$ and up to \unit[40]{Myr} in stellar age. The bias introduced by rapid rotation models does not significantly impact the rest-UV luminosity function or the relationship between UV slope and redshift. This possible bias has a minor influence on the `dust-free' interpretation and brings the sSFR more in line with the predictions from theoretical models. 
\end{itemize}

\begin{acknowledgements}
We thank the anonymous referee for their valuable comments.

\newline
{\it Software:} PARSEC \citep[2.0;][]{2022A&A...665A.126N}, 
	Astropy \citep{2013A&A...558A..33A},
	IPython \citep{2007CSE.....9c..21P},
	Prospector \citep{2017ApJ...837..170L,2019ApJ...876....3L,2021ApJS..254...22J},
    FSPS \citep{2009ApJ...699..486C,2010ApJ...708...58C,2010ApJ...712..833C,2010ascl.soft10043C,2014zndo.....12157F},
    EAZY \citep{2008ApJ...686.1503B},
    Matplotlib \citep{2007CSE.....9...90H}
\end{acknowledgements}

\end{document}